# Robustness and Imperceptibility Enhancement in Watermarked Images by Color Transformation

Maedeh Jamali, Mahnoosh Bagheri, Nader Karimi, Shadrokh Samavi
*Isfahan University of Technology*
*Isfahan, 84156-83111 Iran*

*Abstract*—One of the effective methods for the preservation of copyright ownership of digital media is watermarking. Different watermarking techniques try to set a tradeoff between robustness and transparency of the process. In this research work, we have used color space conversion and frequency transform to achieve high robustness and transparency. Due to the distribution of image information in the RGB domain, we use YUV color space, which concentrates the visual information in the Y channel. Embedding of the watermark is performed in the DCT coefficients of the specific wavelet subbands. Experimental results show high transparency and robustness of the proposed method.

## I. INTRODUCTION

Watermarking is a protection technique that is used to decrease the concern of digital data copyright, such as video, audio, and images. Image watermarking means hide an image into another image. They named the watermark image and host image, respectively.

Watermarking methods can be classified into several categories. One of them is blind or non-blind. That means whether or not additional information is needed for watermarking. The proposed method in [1], falls into a non-blind category because for extract watermark image from a watermarked image, the host image is needed. But in [2,3] the methods are blind and can extract watermark images without the need for the host image.

The other category is the domain that the watermark image uses for the embedding process. There are two domains: spatial and transform. Spatial domain methods are usually less susceptible to conventional image processing attacks because the information is directly embedded in image pixels and easily manipulated. There are various techniques in the transform domain such as Discrete Cosine Transform (DCT) [3]-[5], Fourier Discrete Transform (DFT), Discrete Wavelet Transform (DWT) [6]-[8], Contourlet Transform (CT) and Hadamard Conversion [2]. In [3], authors introduce an adaptive blind watermarking in which the watermark is embedded in DCT coefficients of CT. Two-level CT is applied to the host image. They divided the approximate image into blocks in first levels. Then they extracting the important edges of each block using their proposed edge detection method. These areas are candidate regions for strong embedding. Some parts of the second level are also concatenating with the mentioned blocks. The entropy of blocks and some other image metrics of each block produce an adaptive strength factor for that block. Finally, the DCT transform of blocks is used in their embedding algorithm. In [4], Fang et al. find the relationship between positions and the magnitude of changes in the DCT coefficient and direction of tissue blocks. The direction factor mapping designed by examining such a relationship. The method proposed in [6] is a non-blind watermarking. It uses the HL sub-band of the first DWT of 512×512 image to embed a watermark image with a size 32×32. A geometric algorithm is proposed for embedding to generate a trade-off between robustness and imperceptibility in [8]. They used eight samples of wavelet approximation coefficients from each image block and built two line-segments in a two-dimensional space. Some methods only work in one domain, and others work in a combination of domains. Some approaches use multi transforms. For example in [9], DCT and DWT transforms are used together. The proposed method is a combination of DCT, DWT and fuzzy system for embedding. They first transformed the image into DWT for two levels and selected the LL region in both transformations. Then divide a selected area into 8×8 blocks and calculate the DCT of them. Three attributes related to HSV are fed to a fuzzy system to calculate the strength factor for each DCT block. After that, they manipulate the DCT coefficients based on these adaptive strength factors and watermark bits.

Some papers use color image watermarking space conversion to increase robustness. There are several color spaces like RGB, HSI, La*b*, YCbCr. Usually, authors prefer to use YCbCr color space because the Human Visual System (HVS) is more sensitive to luminance than color. In RGB color space, all colors have the same resolution, but in YCbCr, Y has high resolution, but Cb and Cr have lower. In [5], the host and cover images are color images. The authors convert the color watermark image to binary image, and for the host image, convert it to YCbCr and use channel Y for embedding. The channel Y is divided into 8×8 blocks and after apply DCT to each block, the (3,1) and (2,2) DCT coefficients are used to embed a binary watermark. Roy et al. [10], also convert the host image to YCbCr color space, but they insert the watermark image into the Cb channel. Four-level DWT applied on this channel, and after that, the HL sub-band select. The SVD applies to host and watermark image and the linear combination of their singular values and calculated strength factor replaced by the singular value of HL sub-band. The color space used in [7] is Uncorrelated Color Space (UCS). In this paper, both input images convert to UCS and apply three-level DWT on each of them. To improve the quality and robustness, the artificial bee colony used for optimizing the strength factors.

In this paper, a redundancy based watermarking method in color space is proposed. We use a reversible integer to integer transform from RGB to YUV that helps us to convert these color spaces to each other without missing information [11]. The selection of YUV instead of RGB for embedding is because the intensity information in this color space is separate from color information and we have a robust embedding with good visual quality. After converting from RGB to YUV, we decompose each color channel to the wavelet domain. HL and LH of Y, LL of U, and LL of V sub-bands of wavelet domains are selected for embedding. Then these sub-bands are divided into 8×8 blocks and are DCT transformed. The embedding is done by manipulating some DCT coefficients. In the extraction phase, due to redundant embedding, we use voting to calculate the final watermark logo. Our method provides a good tradeoff between transparency and robustness, and it has considerable results against different attacks. It is a blind watermarking method that does not need extra information during extraction.

The rest of the paper is organized in the following manner: In section II, we present our method. Experimental results are presented in Section III, and Section IV is dedicated to concluding remarks.

## II. PROPOSED METHOD

Our proposed method is explained in this section. As mentioned before, both the imperceptibility and robustness are essential in watermarking methods. Hence, we used color images, which in comparison to gray images, have more space for embedding due to their more channels. By having more space for embedding, we can have redundant embedding that will improve robustness in our method. A reversible transform into YUV color space is used that we can be sure we will not miss information during transformation. This transformation and its inverse are described in (1) [11]:

$$\begin{cases} Y = \left\lfloor \frac{R + 2G + B}{4} \right\rfloor \\ U = R - G \\ V = B - G \end{cases} \leftrightarrow \begin{cases} G = Y - \left\lfloor \frac{U + V}{4} \right\rfloor \\ R = U + G \\ B = V + G \end{cases} \quad (1)$$

where, $R$, $G$ and $B$ are the channels of RGB color space. The Y channel contains visual information that by weak embedding in this channel and strong embedding in U and V, we have kept transparency and robustness together. Because some attacks such as JPEG try to change Y channel less than other channels and also Y has a powerful effect on imperceptibility, so selecting this channel beside two other channels help us to have a robust algorithm. The DCT transform of the blocks of the wavelet channels is applied for embedding. The method is described in detail in the following sections. Section A contains the embedding method, and the extraction algorithm is illustrated in section B.

### A. Embedding scheme

In the embedding stage, first, the RGB color image is transformed into YUV color space. Unlike RGB, the luminance information of YUV is kept in Y channel that will change less than two other channels by some attacks such as JPEG. Each channel is used as a cover image for embedding the watermark. After transforming each channel into a wavelet domain, the horizontal (HL) and vertical (LH) coefficients of Y and approximation (LL) coefficients of U and V are DCT transformed respectively. Approximation coefficients of DWT are robust against different attacks but they are also sensitive to changes, and alternating them can decrease the transparency. Therefore, we select HL and LH channels of DWT for Y to have fewer visual effects on the output image. Also, we need to have the robustness against attacks for other channels, so the LL sub-band is chosen as candidate space for embedding in U and V. Finally, the combination of these selected areas for embedding helps us to have a robust and imperceptible method.

The watermark logo is a $32 \times 32$ binary image that is inserted in all three channels in a specific manner. Embedding in all channels helps us to have a pooling between extracted watermarks and improve the robustness. Hiding watermark bits in DCT is done by manipulating DCT coefficients based on the value of watermark's bit. The details of the embedding algorithm are summarized in the following steps. Watermark logo and color image are the inputs of it, and the RGB watermarked image is the output of the algorithm.

The embedding algorithm contains the following steps:

1. Transforming the RGB input image into YUV color space. This transformation is a reversible conversion that ensures us to avoid missing data during this conversion. Also, putting the luminance information on the Y channel instead of all channels helps us to have powerful embedding in two other channels and using the Y channel for helping to extraction during voting.

2. Decomposing each channel into four scales using wavelet transform. We use different wavelet channels for embedding. Because the Y channel includes intensity information and rough manipulation can reduce the visual quality, so we select HL and LH coefficients for embedding. We also use Y channel because JPEG attacks change it less than two other channels, and so during voting can help us. The LL channel of U and V are selected in this stage as a hiding area. The LL channel is robust against different attacks but because they are selected from U and V have less effect on transparency too and help our method for robust embedding against attacks.

3. Partitioning the mentioned wavelet sub-bands into $8 \times 8$ non-overlap blocks. Each color channel's size is $512 \times 512$, and the decomposed image in the wavelet domain has sub-bands with a size of $256 \times 256$, so after partitioning, we will have $32 \times 32$ blocks that have the same size as watermark logo. Four sub-bands are used for embedding.

4. Calculating the DCT coefficients of blocks. The DCT transformation of each achieved $8 \times 8$ block in the previous stage is calculated in this phase. Each block will hide one bit of the watermark logo.

5. Embedding watermark bits in DCT block. The $32 \times 32$ watermark will be embedded in DCT blocks of DWT by

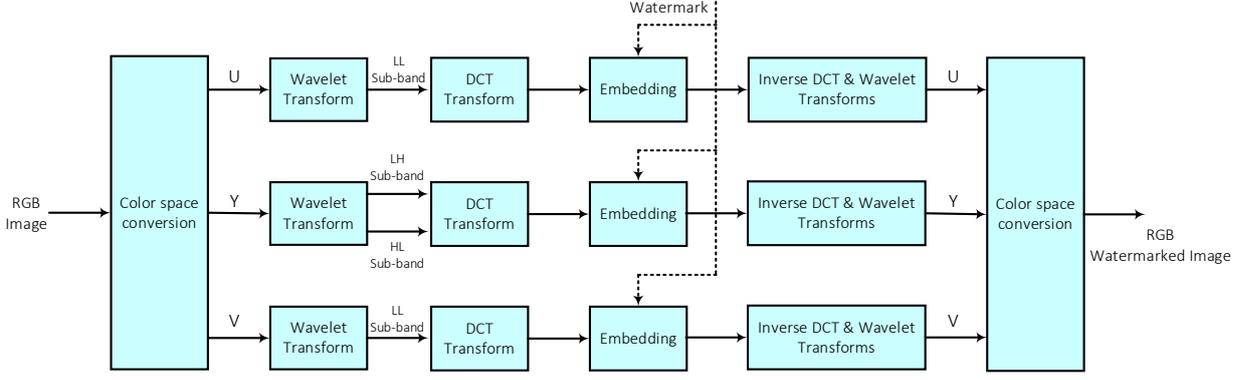

Fig.1. Block diagram of the embedding method.

changing the DCT coefficient. Two DCT coefficients $DCT(u,v)$ and $DCT(v,u)$ of each DCT block are selected that they are placed in reference to the main diameter of the block. The value of them is compared to each other. We consider a rule for embedding that based on this rule and the value of watermark bit, we swap them. To be sure, this change of coefficients will remain after attacks; a strength factor is calculated using (2):

$$\alpha_{i,j} = |DCT_{i,j}(u,v)| + |DCT_{i,j}(v,u)| \times \beta \quad i,j = 1 \text{ to } 32 \quad (2)$$

where $dct_{i,j}(u,v)$ is the coefficient of the block $(i,j)$, $\beta$ is different for each color channel. We use a smaller value for Y than two other channels to keep the transparency. || calculates the absolute value. This factor will increase the difference of two DCT coefficients. The rule in (3) is used for embedding and change the value of coefficients. The coefficients will swap if (3) is not satisfied:

$$\begin{cases} DCT_{i,j}(u,v) - DCT_{i,j}(v,u) \geq \alpha_{i,j} & if\ w(i,j) = 1 \\ DCT_{i,j}(v,u) - DCT_{i,j}(u,v) > \alpha_{i,j} & if\ w(i,j) = 0 \end{cases} \quad (3)$$

6. Repeating the last steps for all other blocks and for different channels. The DCT blocks of two sub-bands of Y and LL sub-bands of U and V will alter based on (3).

7. Calculating the inverse DCT of blocks and inverse DWT of sub-bands. After repeating step 5 for all blocks, the inverse transformation of blocks and sub-bands will be applied, and the watermarked image in YUV will be achieved.

8. Transforming from YUV to RGB. The last step of the embedding method transforms the watermarked image into the RGB color space.

By embedding in two sub-bands of Y and using U and V, we have redundancy in embedding that will improve the robustness of our algorithm against different attacks. The block diagram of the embedding part is shown in fig.1.

*B. Extraction scheme*

Our method is classified as blind watermarking that means we do not send extra information for the extraction process. The receiver can extract watermark by knowing the extraction algorithm. The extraction phase is done as the inverse manner of embedding. In the first step, the watermarked image will be transformed from RGB into YUV. Then Y, U, and V will DWT transformed, and the HL and LH sub-bands of Y, LL sub-band of U, and V will be partitioned into $8 \times 8$ blocks and are DCT transformed.

Each block contains one bit of the logo. Watermark bit extraction will be performed based on the comparison of the value of two DCT coefficients. The extraction from each sub-bands is summarized in (4):

$$w_{i,j-k} = \begin{cases} 1 & if\ DCT(u,v) \geq DCT(v,u) \quad i,j = 1\ to\ 32 \\ 0 & o.w. \end{cases} \quad (4)$$

where $w_{i,j-k}$ indicates the watermark bit in $(i,j)$ location. $k$ means the related sub-bands that are HL, LH of Y or LL of U or V. finally, we will have four watermark maps, and voting should be done to achieve the final watermark. This voting is done based on (5):

$$w\_final_{i,j} = \begin{cases} 1 & if\ \sum_k w_k > 1 \quad k = ll_U, ll_V, hl_Y, lh_y \\ 0 & if\ \sum_k w_k < 2 \end{cases} \quad (5)$$

where $w\_final_{i,j}$ is the extracted watermark after attacks. Figure. 2 shows the block diagram of the extraction algorithm.

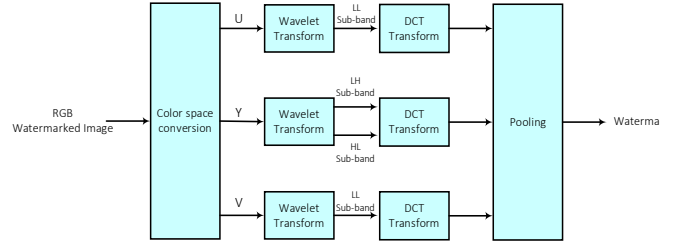

Fig.2. Block diagram of the extraction method.

III. EXPERIMENTAL RESULTS

In order to evaluate our proposed method, several experiments have been considered. As we know, the robustness and imperceptibility act against each other, and increasing one causes the other to decrease. Hence, both of these features should be considered in watermarking methods. Therefore, we report the visual quality and robustness of our proposed method using the Kodak dataset. This dataset contains 24 color images with different sizes, so we resize all of them to $512 \times 512$. The binary watermark logo has a size of $32 \times 32$. The Daubechies (db1) wavelet transform is used for decompositions in three color channels separately for one level. Coordinate of DCT coefficients in DCT blocks are $(v,u) = (6,5)$ and $(u,v) = (5,6)$. These coordinates have good stability against attacks. The RGB to YUV color space is a lossy transformation, and so it can affect the accuracy of the extracted watermark. To evaluate our method, we compare it with a similar embedding and extraction method in RGB color space.

### A. Visual quality

In order to estimate the perceptual quality of watermarked images, different metrics are used. Peak Signal-to-Noise Ratio (PSNR) is a common metric that is reported in watermark papers, but because it ignores the human visual system it is not an accurate metric. Therefore, we also used Structure Similarity Systems (SSIM) in our method. Figure. 3 shows the cover and watermarked image with PSNR and SSIM using our method. We calculate the average PSNR and SSIM on Kodak dataset, and they are 40.6545 and 0.9925, respectively.

### B. Robustness

We evaluate our method against different image processing attacks, such as insertion noise and compression attacks. These attacks include JPEG, salt& pepper (S&P), median filter (MF), Gaussian Noise (GN), Gaussian filter (GF), and sharpening (Sh). Some attacks affect high frequencies, and some of them alter the low frequencies. Therefore, by testing our method against different attacks we can evaluate its robustness thoroughly. Different metrics are applied for robustness evaluation. We use the bit error rate (BER) and Normalize Correlation (NC) here. The NC is calculated based on (6):

$$NC = \frac{1}{m \times n} \sum_{i=1}^{m} \sum_{j=1}^{n} w(i,j) \times \bar{w}(i,j) \quad m,n = 32 \quad (6)$$

where $w$ and $\bar{w}$ are the main and the extracted watermark. Also, $m$ and $n$ indicate the height and width of the watermark. $NC$ value changes between 0 and 1, and values near to 1 show better results. Several attacks are applied on the Kodak dataset. Table I shows the average BER and NC values for all images in the Kodak database. Based on the average values on KODAK, our method has significant results against different attacks.

TABLE I. NC AND BER VALUES OF EXTRACTED WATERMARK FOR DIFFERENT ATTACKS.

| attacks | | GN | | JC | | MF | S&P | | Sh | GF |
|---|---|---|---|---|---|---|---|---|---|---|
| | | 0.001 | 0.006 | 90 | 100 | | 0.001 | 0.006 | | |
| Average on KODAK | BER | 0.0200 | 0.1381 | 0.1693 | 0.0056 | 0.0465 | 0.0179 | 0.0686 | 0.0059 | 0.0034 |
| | NC | 0.9880 | 0.9925 | 0.9005 | 0.9977 | 0.9945 | 0.9901 | 0.9741 | 0.9925 | 0.9993 |

To show the robustness of our method in YUV, we compare our method in YUV and RGB to demonstrate that using this color space instead of RGB helps us to have better imperceptibility and robustness. Table II shows the BER, SSIM and PSNR values for random selected KODAK images in RGB and YUV color space. Based on the result, we can see using this color space for embedding considerably improves transparency, although in some attacks our BER values are higher than the RGB space if we try to produce images with the same PSNR and SSIM in YUV, our BER value will be strongly better than RGB.

These values mean using YUV due to having a separate channel for the intensity, can produce images with better imperceptibility, and so will produce an unrecognizable watermark image.

TABLE II. COMPARISON THE ROBUSTNESS AND TRANSPARENCY OF OUR METHOD IN YUV AND RGB COLOR SPACE AGAINST DIFFERENT ATTACKS

| Image | Attack | | | | | | | | PSNR | SSIM |
|---|---|---|---|---|---|---|---|---|---|---|
| | BER | | | | | | | | | |
| | GF | GN | | JC | | S&P | | MF | | |
| | | 0.001 | 0.003 | 90 | 100 | 0.001 | 0.003 | | | |
| 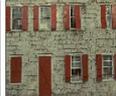 | **0.000** | 0.026 | 0.065 | 0.128 | **0.001** | 0.007 | 0.021 | 0.048 | 40.094 | 0.993 |
| | 0.002 | **0.006** | **0.023** | **0.033** | 0.023 | **0.004** | **0.009** | **0.039** | 27.901 | 0.924 |
| 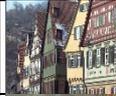 | 0.000 | 0.011 | 0.049 | 0.099 | **0.000** | 0.004 | 0.015 | 0.049 | **38.474** | **0.992** |
| | 0.000 | **0.004** | **0.016** | **0.026** | 0.023 | **0.004** | **0.005** | **0.036** | 26.670 | 0.928 |
| 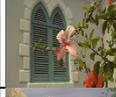 | **0.001** | 0.069 | 0.124 | 0.195 | **0.004** | 0.023 | 0.063 | **0.033** | 41.372 | 0.995 |
| | 0.008 | **0.048** | **0.096** | **0.032** | 0.032 | **0.019** | **0.036** | 0.041 | 29.94 | 0.9613 |
| 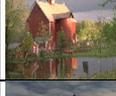 | **0.000** | 0.019 | 0.055 | 0.129 | **0.000** | 0.007 | 0.023 | 0.051 | **40.372** | **0.992** |
| | 0.002 | 0.019 | **0.038** | **0.036** | 0.014 | **0.007** | **0.014** | 0.0469 | 30.369 | 0.951 |
| 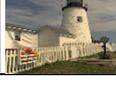 | **0.002** | 0.037 | 0.106 | 0.209 | **0.003** | 0.012 | 0.031 | 0.042 | **42.713** | **0.995** |
| | 0.006 | **0.035** | **0.0527** | **0.044** | 0.026 | **0.007** | **0.0127** | **0.0391** | 30.969 | 0.961 |

In [13] they used color images for embedding using QR decomposition. The host image and watermark logo have the same size as ours, but they use a color logo for embedding. During embedding they change the color logo into binary logo and embed this binary logo into the host image, so we can compare our PSNR and SSIM on using images with them. Table III shows the average mentioned values on same images as [13] (Lena, Jetplane, Peppers, and TTU). Based on Table III we have better transparency with random binary $32 \times 32$ watermark. Methods in [2] and [5] also used color space for embedding but they have different watermark size.

TABLE III. THE AVERAGE PSNR, SSIM AND NC VALUES FOR A BINARY LOGO IN COMPARISON TO [13].

| | PSNR | | SSIM | | NC | |
|---|---|---|---|---|---|---|
| | Ours | [13] | Ours | [13] | Ours | [13] |
| Average on selected images | **37.0516** | 36.6763 | **0.9892** | 0.9367 | 1.0000 | 1.000 |

### IV. CONCLUSION

In this paper, we used color images for embedding that help us to have more space for embedding, and we can use these extra spaces as three channels for redundant embedding. A reversible color space transformation is used in this method and we can embed it in different channels of new space without missing information during this transformation. The embedding is done by manipulating the DCT coefficients of wavelet sub-bands of color channels and by redundant embedding we propose a robust watermark method. By comparing the PSNR, SSIM and BER values of our approach in RGB and YUV color, we demonstrate that YUV color can have better results in transparency with reasonable BER.

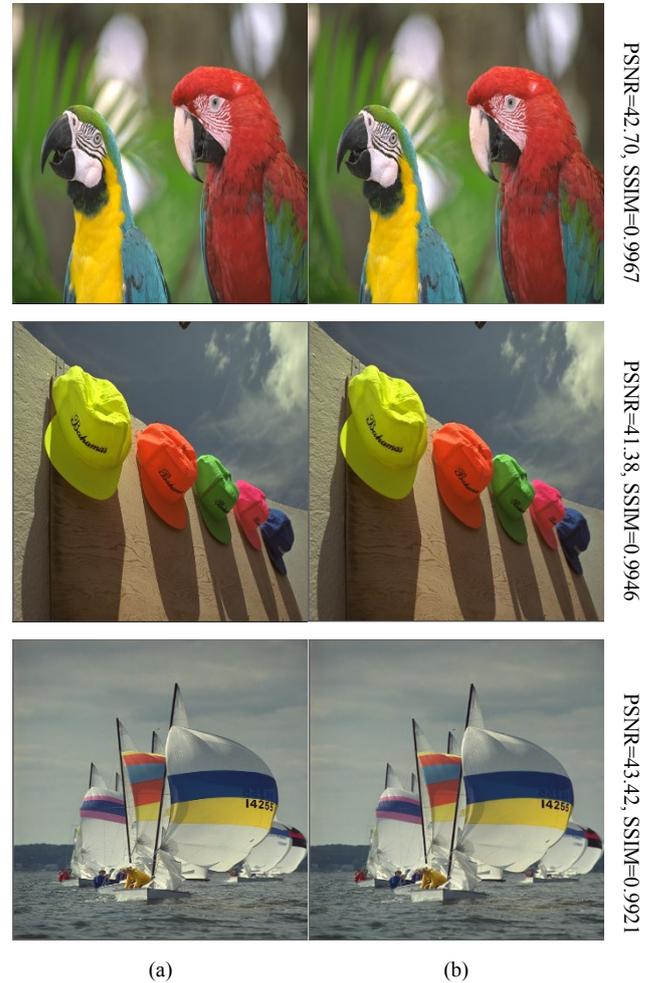

(a)                  (b)

Fig.3 Visual quality comparison of the original image and watermarked image, (a) original image, (b) watermarked image.

PSNR=42.70, SSIM=0.9967

PSNR=41.38, SSIM=0.9946

PSNR=43.42, SSIM=0.9921